\documentclass[a4paper]{article}
\usepackage[latin1]{inputenc} %
\usepackage[T1]{fontenc} %
\usepackage{RRA4}
\usepackage{amsmath,amssymb}
\usepackage{subfigure}
\usepackage{hyperref}
\RRdate{September 2009}

\RRauthor{
Nidhi Hegde\thanks[orange]{Orange Labs}%
  \and Fabien Mathieu\thanksref{orange}
  \and
Diego Perino\thanks{INRIA} \thanksref{orange}%
}
\authorhead{Hegde, Mathieu \& Perino}
\RRtitle{Diffusion épidémique par chunks : la taille compte}
\RRetitle{Size Does Matter\\ (in P2P Live Streaming)}
\titlehead{Size Does Matter in P2P Live Streaming}
\RRresume{Le problème de la diffusion pair-à-pair en quasi-direct consiste à transmettre un contenu à un ensemble de pairs avec la meilleure qualité possible tout en minimisant le différé, c'est-à-dire le délai d'acheminement de la source aux pairs. Dans un grand nombre de solutions, le contenu est découpé en quantités de taille fixée, les \emph{chunks}. En supposant que le temps de transfert d'un chunk d'un pair à un autre est uniquement déterminé par la bande passante de l'émetteur, le délai optimal possible est typiquement en $\frac{c}{s}(\log_2(n))$, $c$ étant la taille des chunks, $s$ le débit du contenu diffusé et $n$ le nombre de pairs. Il semble alors naturel de choisir $c$ aussi petit que possible afin de minimiser le délai. Il arrive cependant un point où les latences présentes dans le réseau influent nécessairement sur la diffusion du contenu.

Notre objectif est de mettre en évidence les phénomènes qui apparaissent lorsque la taille du chunk rend les effets de latence non négligeables. En se basant sur un mécanisme de diffusion épidémique simple, nous mettons en évidence les effets suivants :
\begin{itemize}
	\item des chunks trop petits empêchent l'algorithme de fonctionner efficacement, et génèrent un taux de pertes important ainsi qu'un gaspillage des ressources réseau ;
	\item à partir d'une certaine taille, les pertes cessent et la quantité de messages de contrôle se stabilise, le délai étant proportionnel à la taille du chunk ;
	\item entre les deux se situe un intervalle de tailles adaptées à la diffusion. Le choix d'une taille précise dépend du compromis à réaliser entre pertes et délai.
\end{itemize}
De plus, nous observons que l'introduction d'un certain parallélisme dans la diffusion permet de déplacer la zone utile, augmentant ainsi la performance de l'algorithme.}

\RRabstract{Optimal dissemination schemes have previously been studied for
peer-to-peer live streaming applications. Live streaming being a delay-sensitive
application, fine tuning of dissemination parameters is crucial. In this paper,
we investigate optimal sizing of chunks, the units of data exchange, and probe
sets, the number peers a given node probes before transmitting chunks. Chunk
size can have significant impact on diffusion rate (chunk miss ratio), diffusion
delay, and overhead. The size of the probe set can also affect these metrics,
primarily through the choices available for chunk dissemination. We perform
extensive simulations on the so-called random-peer, latest-useful dissemination
scheme. Our results show that size does matter, with the optimal size being not
too small in both cases.}
\RRmotcle{Pair-à-pair, diffusion en quasi-direct, délai, taille des chunks}
\RRkeyword{P2P, Live Streaming, Delay, Chunk Size}
\RRprojet{Gang}  
\RRdomaine{1} 
\RRtheme{COM -- Systèmes communicants}
\RRNo{7032}
\URRocq 

\newcommand{\rplu}{\emph{rp/lu} }

\begin{document}
\makeRR   

\tableofcontents
\newpage

\section{Introduction}

Peer-to-peer data transfer has been  a dominant source of network traffic for the past few years.  Peer-to-peer mechanisms of transfer that rely on client uploads have also been  used recently for live video streaming solutions such as PPLive~\cite{pplive}, CoolStreaming~\cite{coolstreaming}.   Natural questions that arise for live video streaming concern whether the delay and quality requirements can be met by distributed client-based dissemination.

In most cases, a P2P live streaming algorithm splits the streams into chunks (also termed pieces).  Chunks are considered as the atomic components of the stream, and a peer can only send chunks it has fully received.  Much of the work in the literature has been devoted to the search for a chunk exchange policy that is feasible and optimal.  We focus here on the unstructured epidemic approaches~\cite{massoulie07randomized,bonald08epidemic,piconni08future}, where the policy is described by a \emph{scheme} that indicates which chunk a given peer should try to send, and to whom. 

A good scheme is indeed essential to the epidemic live streaming problem.  For a given scheme however, an optimization at a detailed level is also important.  This involves the fine tuning of dissemination parameters, such as chunk size, receiver buffer size, number of peers to probe, etc.  The chunk size has a significant impact on performance, since smaller chunk sizes may be more efficient but incur relatively higher overhead, and larger chunk sizes have lower overhead but may result in higher delay.  The receiver buffer size (relative to chunk size) impacts the diversity in choice available to a peer for transmission.  In the scheme with random peer choice, probing more than one peer for the decision of chunk exchange may help (power of choices), but it also increases overhead.  These are some of the finer details of any dissemination scheme that must be closely examined.  

There has been some study on parameter sizing for peer-to-peer file sharing systems.  In~\cite{legout08small} it is shown that small chunk sizes are not always best for file transfer; \cite{bitmax} proposes uplink allocation strategies designed to improve uplink utilization of BitTorrent-like systems. However, results obtained for file sharing systems are not directly applicable to live streaming applications.  First, a newly created chunk should be disseminated as fast as possible in live streaming, so there is a strong delay component, naturally limiting the chunk size.  Secondly, missing chunks may be acceptable if a resilient codec is used, so optimal values are not always comparable to those in the file transfer case.  Then, the buffer size, which is a parameter specific to streaming, can impact the performance (see for instance ~\cite{ZhouCL07}).

In this paper, we investigate dissemination parameters in peer-to-peer live video streaming through extensive simulations.  Specifically, we focus on the \emph{rp/lu} diffusion scheme, where a peer sends the latest (freshest) useful chunk to a randomly selected peer. We will also briefly consider other schemes for comparison.   We will show that indeed chunk size significantly impacts the performance.  In fact, there is a range of chunk sizes that may be suitable, where the specific choice of the chunk size ultimately depends on the delay/chunk miss ratio trade-off.  We will also show that a fine tuning of the number of peers to probe and the number of simultaneous chunks to send is important.  

The rest of the paper is organized as follows.  In the next Section we outline our simulation framework.  Section~\ref{sec:performance}  covers the impact of the chunk size, and highlights the suitable range of chunk sizes among various dissemination schemes.  Section~\ref{sec:probeset} examines the value of the number of peers to probe for chunk dissemination.  We finally conclude the paper in Section~\ref{sec:conclusion}.

\section{Methodology}

Epidemic diffusion schemes and their behavior have been extensively studied in the literature.   See for instance~\cite{bonald08epidemic} for a detailed study and references therein.   Here, we consider schemes where the sender first selects the destination peer and then sends a chunk. 
We focus on this particular class of schemes because they are fairly simple and thus allow us to focus on the impact of the parameters such as chunk size.  Moreover, they are efficient in terms of diffusion rate and delay, and can potentially generate low overhead. 

In particular we consider  three algorithms representative of this class:
\begin{description}
\item[\emph{random peer / latest blind chunk} (\emph{rp/lb}).]
The destination peer is chosen uniformly at random among sender peers' neighbors and the most recent chunk in the buffer is selected (regardless of whether the receiver needs that chunk or not);
\item[\emph{random peer / latest useful chunk} (\emph{rp/lu}).] 
The destination peer is chosen uniformly at random among sender peers' neighbors and the most recent chunk not own by the receiver peer is selected; unless otherwise specified, this is the scheme considered in this paper.
\item[\emph{bandwidth aware peer / latest useful chunk (\emph{ba/lu})}.] 
This scheme is inspired by~\cite{mellia08bandwidth}.
A peer $i$ is selected with a probability proportional to its upload bandwidth $u_i$ and the most recent chunk not own by the receiver peer is selected.  Note that for homogeneous upload bandwidths this is equivalent to \emph{rp/lu}.  
\end{description}

In order to analyze these algorithms under the same framework and derive general results, we used an event-based simulator developed by the Telecommunication Networks Group of Politecnico di Torino\footnote{\url{http://www.napa-wine.eu/cgi-bin/twiki/view/Public/P2PTVSim}}. The simulator has been modified to take network latencies, control overhead and parallel upload connections into account.

In our simulator we assume that the overlay network is an Erd{\"o}s-Renyi graph $\mathcal{G}(n,p)$,  where $n$ is the size of the peer population and $p$ is the probability that a link connecting two peers does exist. Every peer $i$ has therefore a partial view of the overlay network, with an average number of neighbors $p(n-1)$. 

We assume that every link connecting a pair of peers $\{i,j\}$ is characterized by a constant round trip delay $\text{RTT}_{ij}$ and is lossless.  We further assume that there are no queuing nor processing delays, so the \emph{transfer delay} (the time for a chunk or control packet to travel from peer $i$ to peer $j$) is equal to $transmission~delay+\frac{RTT_{ij}}{2}$.
The choice of such a network model allows us to obtain results that are not affected by the overlay network structure or by transport network congestion or losses.
A peer is characterized by its upload bandwidth $u_i$.  There is a single source $S$ with upload capacity $u_s$ and a limited overlay knowledge as well.

Every peer periodically selects a subset $m$ of its neighbors, according to one of the aforementioned algorithms (that is random or bandwidth-aware selection), and probes them in order to discover their missing chunks, except for the case of the \emph{latest blind} scheme.   We refer to the set of neighbors probed as the probe set.  Based on the responses possibly received, the peer then transmits corresponding chunks.

A peer can upload a chunk to at maximum $m'$ peers in parallel by fairly sharing its upload bandwidth.  It may happen that a peer cannot serve $m'$ recipients because it does not have enough useful chunks. In that case it uploads the chunks faster (since there are less than $m'$ active connections), but it may stay idle for the subsequent period of time (because it needs to acquire new chunk maps from newly selected peers).
An additional overhead is taken into account at every peer to reply to control messages coming from potential sender peers.

Unless otherwise stated we consider a network of  $n=1000$ peers, all with the same upload bandwidth $u_i=1.03 Mb/s$, an unlimited download bandwidth and about $50$ neighbors ($p=0.05$).  We set the stream rate $s=0.9 Mb/s$.  Latencies between nodes are taken from the data set of the Meridian project~\cite{meridian}.  A buffer of size up to $300$ chunks is available at all peers, in order to avoid possible missing chunks due to buffer shortage (this implies a buffer size proportional to the chunk size).

\section{Chunk Size and performance}
\label{sec:performance}

When considering a streaming algorithm, a crucial performance metric is the diffusion rate/diffusion delay/overhead trade-off achieved by that algorithm, which can be summarized by a (\emph{chunk miss ratio},\emph{delay},\emph{overhead}) triplet.

Following~\cite{liang08random}, we define the \emph{chunk miss ratio} as the asymptotic probability to miss a chunk (or equivalently the difference between the stream rate $s$ and the actual goodput), while \emph{average diffusion delay} is defined as the delay between the creation of a chunk and its reception by a peer, averaged over the successful chunk transmissions. Note that since links are lossless, a peer misses a given chunk only if none of its neighbors has scheduled that chunk for it.
The \emph{overhead} is  defined as the difference between the bandwidth used by peers (throughput) and the actual data received (goodput).  In our framework, this overhead is due only to control messages exchanged between peers.

As a first experiment, we analyze the performance triplet as a function of the chunk size.  The results are shown in Figures~\ref{fig:homo_rate} to~\ref{fig:overhead}, for the \rplu~scheme with $m=m'$  varying from $1$ to $5$.

\subsection{Chunk miss ratio}
\label{subsec:losses}

\begin{figure}[ht]
\centering
	\includegraphics[width=.60\textwidth]{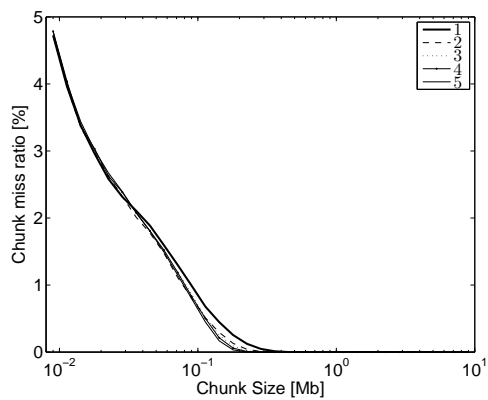}
	\caption{Chunk miss ratio as a function of the chunk size ($m=m'$ varying from 1 to 5).}
	\label{fig:homo_rate}
\end{figure}

In Figure~\ref{fig:homo_rate}, we observe two cases:
\begin{itemize}
	\item For large chunks (in our experiment, $c$ greater than a few hundred kilobits, the exact value depends on the number of simultaneous connections $m$), there are no missing chunks.
	\item As the chunk size goes below a certain critical value, chunks start to miss, roughly proportional to the logarithm of the chunk size.
\end{itemize}

This phenomenon can be explained as follows: the time between two consecutive chunks is $c/s$, and is therefore proportional to the chunk size $c$.  When $c$ is big enough (all other parameters being the same), we can assume that more and more control messages per chunk can be exchanged between peers.  This should achieve a proper diffusion, provided enough bandwidth is available, since  a sender peer will have  enough time to find a neighbor needing a given chunk.  On the contrary, when $c/s$ is too small, peers do not have enough time to exchange control messages, resulting in missing chunks.
Note that increasing $m$ slightly improves the performance.

\subsection{Delay}
\label{subsec:delay}

\begin{figure}[ht]
\centering
	\includegraphics[width=.60\textwidth]{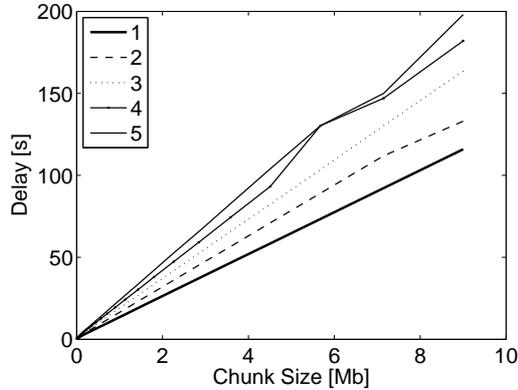}
	\caption{Average diffusion delay as a function of the chunk size}
	\label{fig:homo_delay_mean}
\end{figure}

The average diffusion delay as a function of the chunk size is shown in Figure~\ref{fig:homo_delay_mean}. The main result is that the delay is proportional to the chunk size (hence the linear x-axis used; although difficult to observe on the figure, the proportional relationship was also verified for small values.). We also note that it grows with $m$.

This result is consistent with theoretical results obtained in~\cite{piconni08future} where RTT is neglected and the chunk transmission time is simply considered inversely proportional to the sender's bandwidth.  Under that framework, the minimal diffusion delay is given by: 

\begin{equation}
	d_{\min}=\frac{mc\ln(n)}{\ln(1+m)s}.
	\label{eq:opti_diff_delay}
\end{equation}

\subsection{Overhead}
\label{subsec:overhead}

\begin{figure}[ht]
\centering
	\includegraphics[width=.60\textwidth]{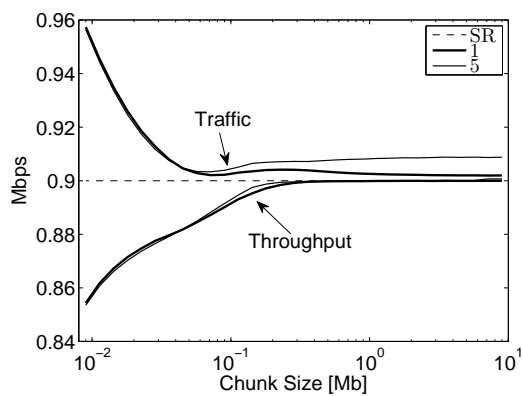}
	\caption{Goodput and throughput as a function of the chunk size, the overhead being the difference. The stream rate $s$ is also indicated.}
	\label{fig:overhead}
\end{figure}

The performance with respect to overhead, i.e. the difference between the throughput and goodput, is shown in Figure~\ref{fig:overhead} (only the curves for $m=1$ and $m=5$ are displayed for legibility).    For very small chunks, we have a non-intuitive trend, where as $c$ grows, the goodput increases \emph{and} the throughput decreases (or equivalently, the overhead decreases faster than the goodput increases).  This process slows down so that at some point the throughput increases again.  For big enough chunks, the overhead becomes roughly constant (for a given $m$), while the goodput becomes equal to the stream rate (meaning no missing chunks).

The goal of this paper is not to give a complete explanation of the observed results, but rather give some intuitions behind the overhead behavior.  For very small chunks, chunk miss ratio is high, which, as mentioned earlier, come from the fact that not enough control messages can be sent.  Asymptotically, we may imagine that only one control message per sent chunk is produced, resulting in an overhead/goodput ratio of $\frac{c_c}{c}$, where $c_c$ is the size of a control message.

On the other hand, in the limit as the chunk size is increased, we may expect that a peer can send a number of messages per sent chunk that is proportional to the chunk characteristic time $c/s$.  This would result in an overhead ratio proportional to $\frac{c_c}{s}$, and thus independent of $c$ (but not of other parameters like the median RTT or $m$).

\subsection{\texorpdfstring{Suitable range for $c$}{Suitable range for \emph{c}}}
\label{subsec:suitable}

In light of the study above, there is a good order of magnitude for suitable chunk size in epidemic live streaming. For the parameters considered here, $c$ should be greater than $0.06$~Mb (which corresponds to about $15$ chunks per second) and smaller than $0.3$~Mb ($3$ chunks per second):
\begin{itemize}
	\item to send the stream at more than $15$ chunks per second is good for the delay (which stays roughly proportional to $c$), but results in both an increase in throughput and a decrease in goodput;
	\item goodput and throughput are stationary for $c$ greater than $0.3$~Mb:  using bigger chunks only means longer delay;
	\item between these values, the choice of $c$ results in a chunk miss ratio/delay trade-off: smaller delay with some missing chunks or greater delay with no missing chunks.  Choosing a precise value for $c$ depends then on factors that will not be discussed here, such as the codec used, the required QoS, etc.
\end{itemize}

\begin{figure}[ht]
\centering
	\includegraphics[width=.60\textwidth]{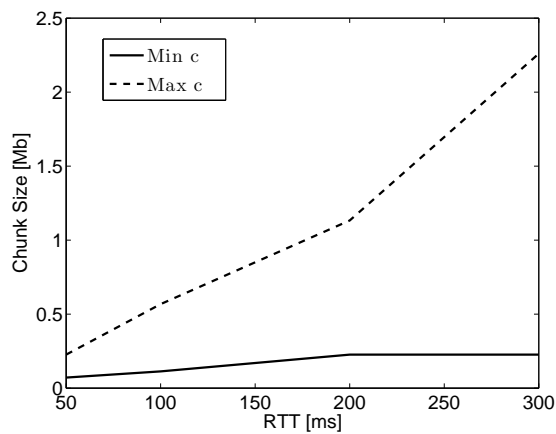}
	\caption{Suitable range (for $m'=m$)}
	\label{fig:useful}
\end{figure}

In our experiments the suitable range for chunk size begins when the chunk characteristic time ($\frac{c}{s}$) has the same order of magnitude than the median RTT, and ends an order of magnitude later.  We scaled the RTT distribution used in order to observe the evolution of the range with the median RTT. The results, reported in Figure~\ref{fig:useful}, show that the range values are indeed roughly proportional to the median RTT.

Note that the lower bound of the suitable range gives an indication on the minimal delay that can be achieved without too much missing chunks and overhead.   In section~\ref{sec:probeset}, we will see that enhanced diffusion techniques can help lower that bound.

We have performed experiments using various diffusion schemes, RTT and bandwidth distributions, values of probe set $m$, stream rate $s$ and so on.  All results are not report here for lack of space but, even if given metric values may differ, we observed the existence of a \emph{suitable} range for $c$.

\begin{figure}[ht]
\centering
\subfigure[Average goodput and throughput]{
	\includegraphics[width=.5\textwidth]{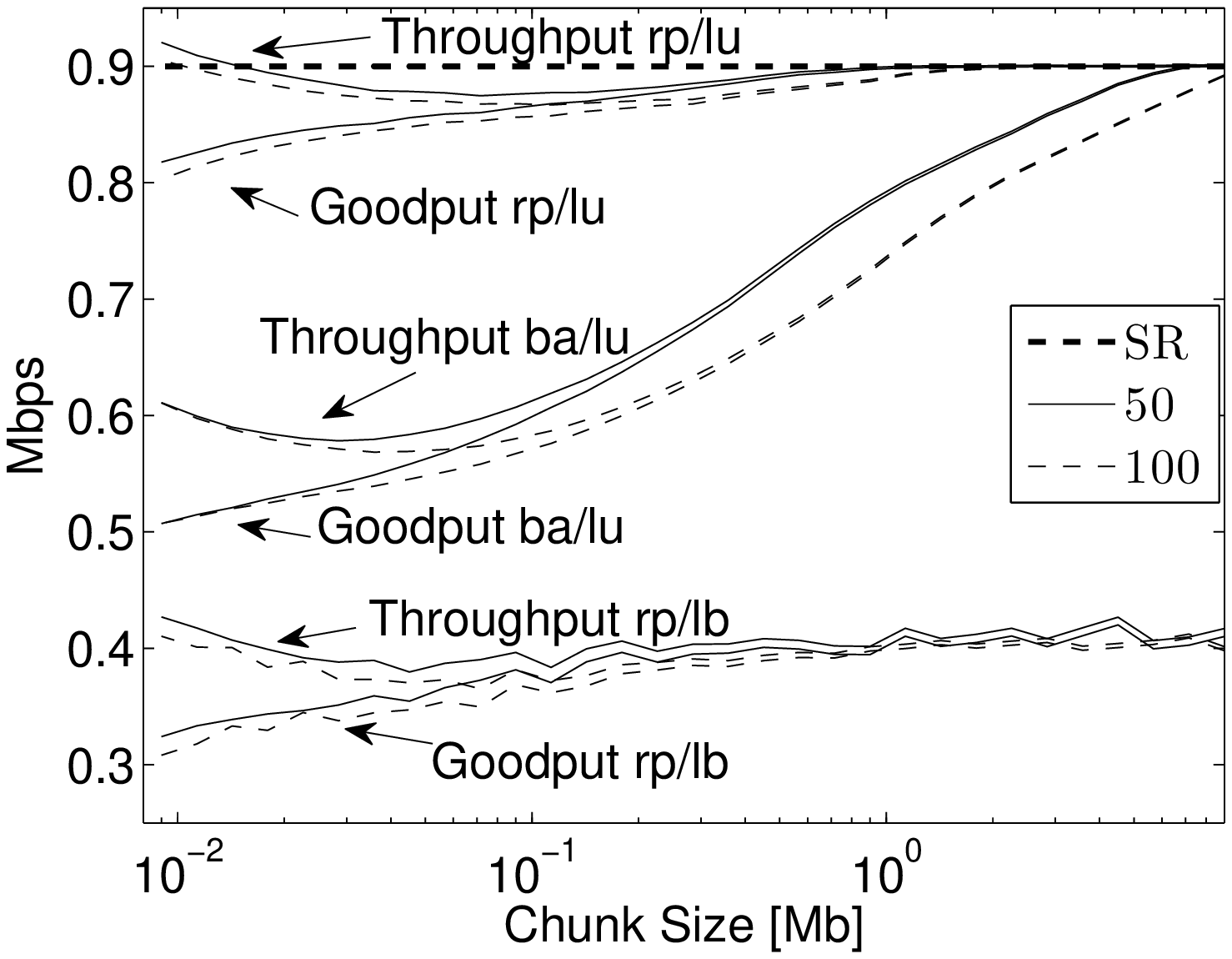}
	\label{fig:comparison_rate}}
\subfigure[Average diffusion delay]{
	\includegraphics[width=.45\textwidth]{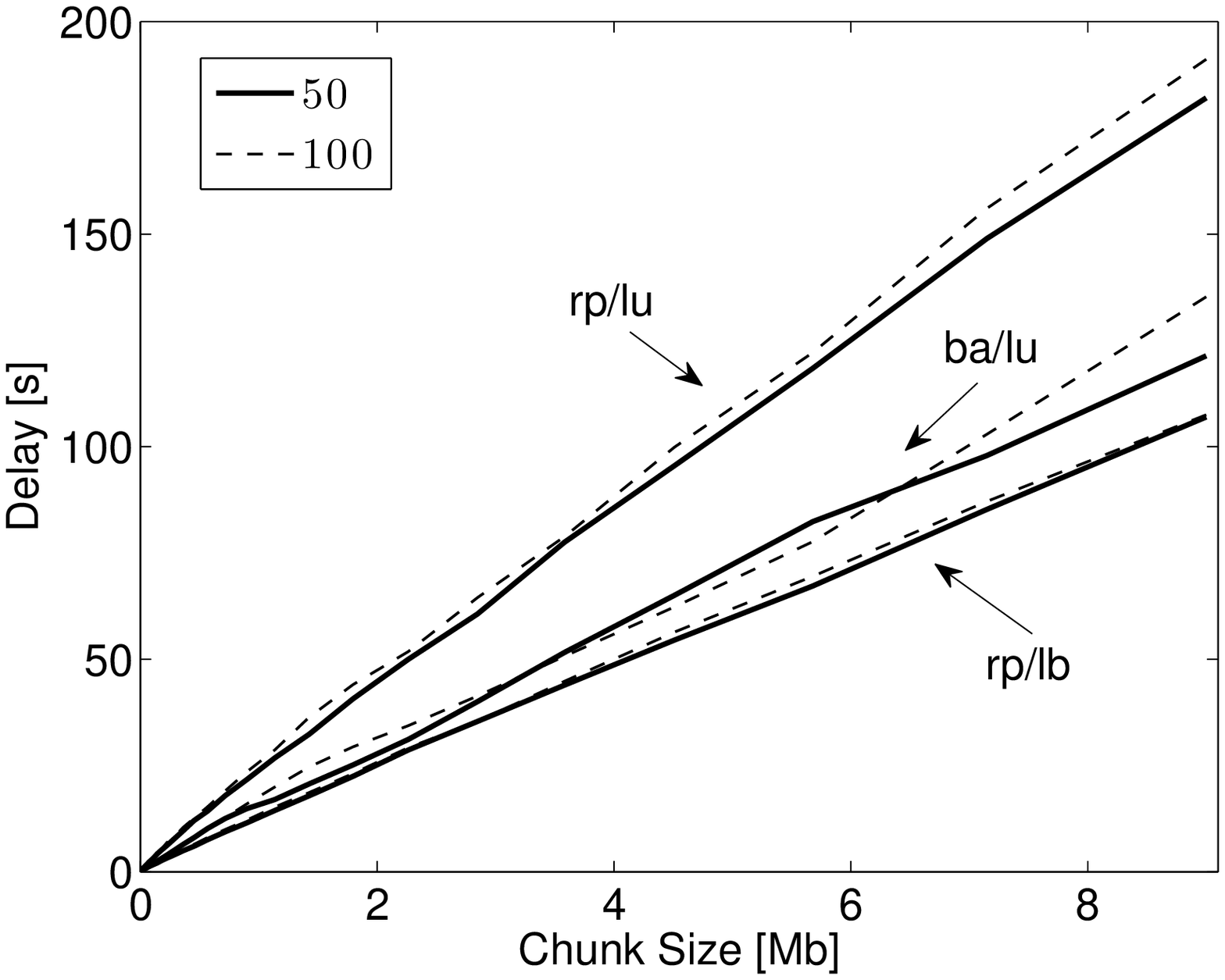}
	\label{fig:comparison_delay_mean}}
	\caption{\emph{rp/lb}, \emph{rp/lu} and \emph{ba/lu} comparison}
	\label{fig:comparison}
\end{figure}

As an example, in the following we compare the suitable range of chunk size for two RTT values and the three following dissemination schemes: \emph{rp/lu}, \emph{rp/lb} and \emph{ba/lu}. Since the scenarios with homogeneous bandwidths are identical under the \emph{rp/lu} and \emph{ba/lu} schemes, we use a heterogeneous bandwidth distribution derived from~\cite{saroiu02measurement}.   We set $m=m'=1$, and we plot the throughput, goodput and average delay for these cases using two values of latency, $\text{RTT}=50, 100$~ms (Figure~\ref{fig:comparison}).  

Note that the scheme \emph{rp/lb} suffers high chunk miss ratios for all values of chunk sizes considered.  Indeed it has been shown~\cite{bonald08epidemic} that this scheme performs poorly with respect to rate, while being optimal with respect to delay.   The scheme \emph{rp/lu} has fewer missing chunks, but higher delay, while the performance of \emph{ba/lu} lies between the other schemes for both chunk miss ratio and delay.

However, beyond the fact that the chunk miss ratio/delay/overhead trade-off is closely related to the scheme (more complete studies are available elsewhere~\cite{bonald08epidemic,piconni08future,mellia08bandwidth}), the striking observation is that all these schemes admit a similar suitable range for $c$, which seems to scale with the median RTT of the network.  This supports our claim that the suitable range for $c$ depends mainly on the median RTT and $s$ (the inter-chunk delay $\frac{c}{s}$ should have roughly the same order than the RTT), the actual scheme being secondary.

\section{Size of Probe set}
\label{sec:probeset}

In the results presented so far, we have assumed that the number of simultaneous exchange chunks, $m'$, is identical to the size of the probe set $m$.  We now consider the impact of probing more peers than the number of simultaneous chunks sent.  A larger probe set affords a sender peer a higher chance to find a recipient peer for whom it has useful chunks (power of choices principle).   However, it also increases overhead, and possibly delay.

\begin{figure}[ht]
\centering
\subfigure[$c=0.15$ Mb (middle of the suitable range for $m=m'$)]{\includegraphics[width=.45\textwidth]{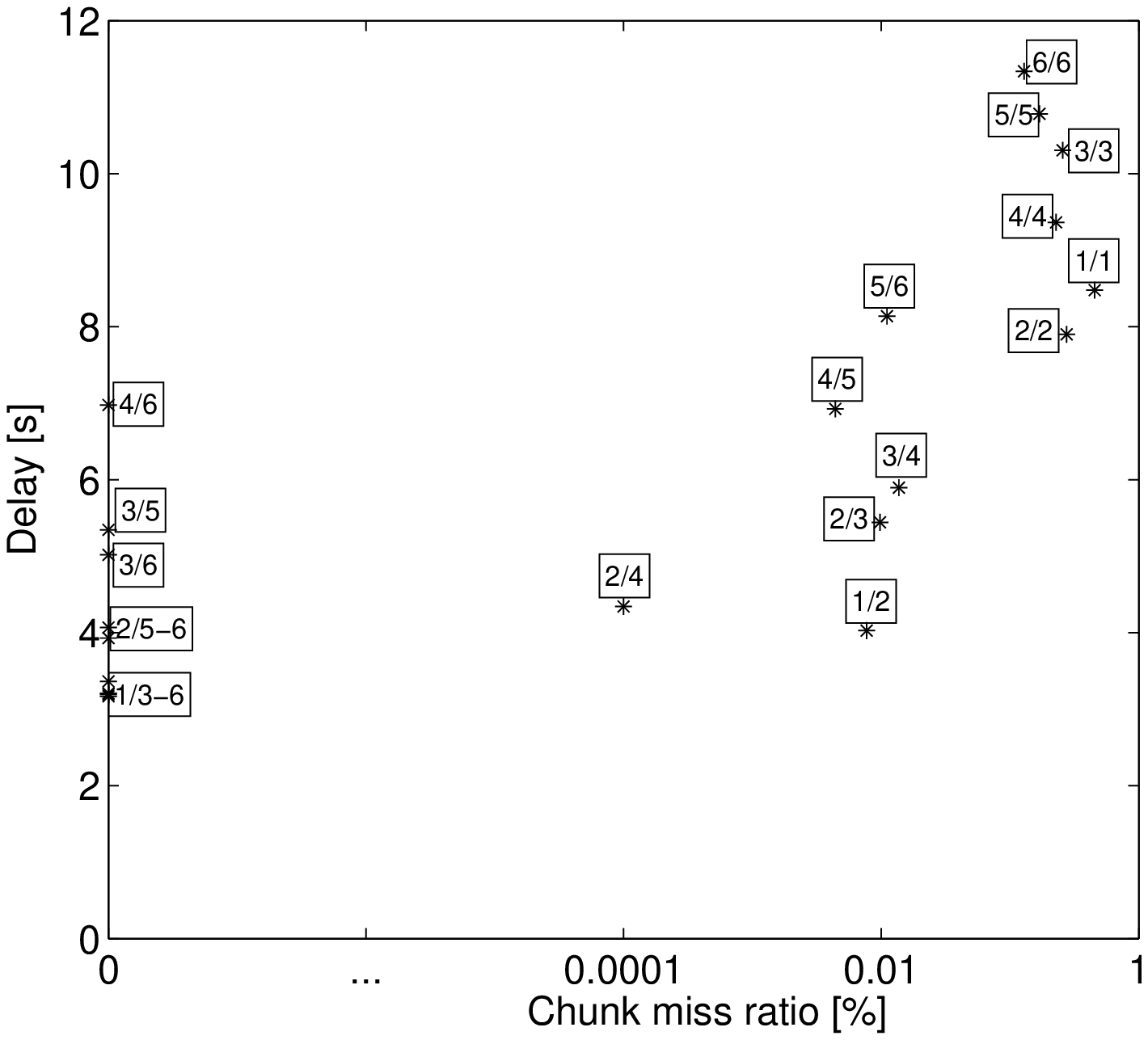}
	\label{fig:CC0}}
\subfigure[$c=0.035$ Mb (below the suitable range for $m=m'$)]{\includegraphics[width=.5\textwidth]{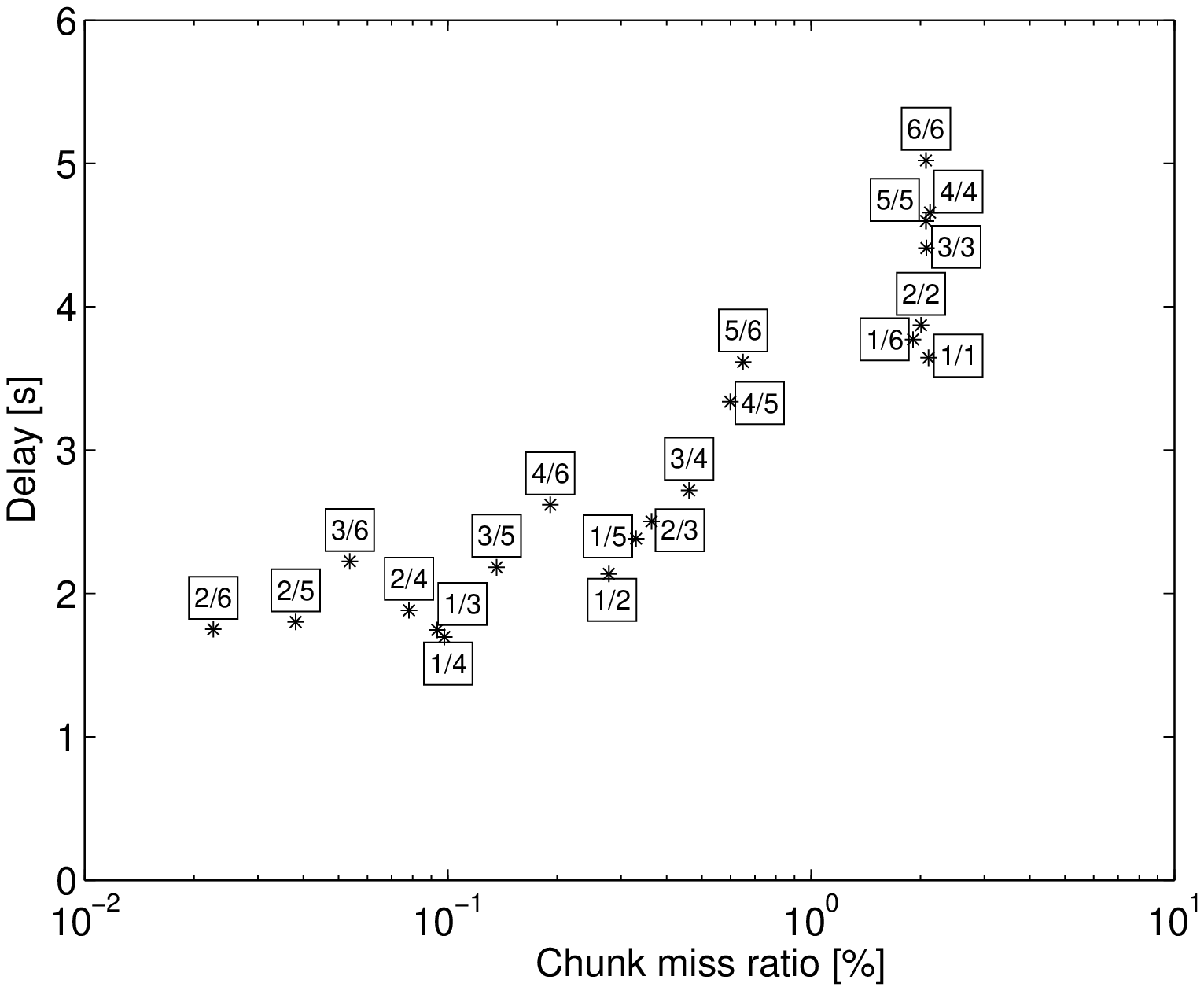}
	\label{fig:CC}}
\caption{$m'/m$ chunk miss ratio/delay trade-off for two values of $c$}
\end{figure}

Figure~\ref{fig:CC0} plots the chunk miss ratio/delay trade-off for various $m'/m$ pairs. The scheme is \emph{rp/lu}, the bandwidth is homogeneous and the chunk size is set to $c=0.15$ Mb (middle of the suitable range).
The figure shows that using $m'=m$ is not optimal, and having a larger probe set, $m>m'$ significantly reduces both delay and missing chunks. 
The delay decreases from about $10$~s for the $m=m'$ case,  to less than $4$~s for the $1/3,\ldots,6$ cases (meaning $m'=1$ and $m=3, \ldots,6$).
With regards to the chunk miss ratio, there are some $(m'/m)$ pairs for which no missing chunks could be observed in our experiment: $1/3-6$, $2/5-6$, $3/5-6$, $4/6$. This suggests that a consequence of using $m'<m$ is a shift of the suitable range for $c$.

In order to verify this interpretation, we now set $c=0.035$ Mb, which is clearly below the suitable range observed in \S~\ref{subsec:suitable} for $m=m'$.  The results are shown in Figure~\ref{fig:CC}.

We observe that no pair $(m'/m)$ can achieve diffusion without missing chunks for such a small $c$, however the trade-offs are still worthwhile with respect to the delay: using $m'/m=2/6$, we get a delay of $1.7$~s with a chunk miss ratio of about $0.02~\%$. This indicates that $c=0.035$ Mb is definitively within the suitable range for $m'/m=2/6$.

Also note how the relative efficiency of the various $m'/m$ values is impacted by the choice of $c$: for instance, $1/6$, which is optimal for $c=0.15$ Mb, performs rather poorly for $c=0.035$ Mb.  Although the results presented here refer to the \emph{rp/lu} scheme, we performed experiments with other schemes and we observed similar trends, confirming that using a proper $m'<m$ can significantly improve the delay.

On the other hand, there is a price for going below the suitable range defined in \S~\ref{subsec:suitable}: for a given scheme, the overhead still depends on $m$ and $c$.  For \emph{rp/lu}, it stays close to the overhead displayed in Figure~\ref{fig:overhead} even for $m'<m$. So using small $c$ with $m'<m$ can reduce the delay, but it requires more throughput.

\section {Conclusion}
\label{sec:conclusion}

We have investigated the dissemination parameters of peer-to-peer epidemic live video streaming through extensive simulations.  We have shown that the chunk size significantly impacts performance and that the chunk size should fall within a given range which is mostly determined by the median RTT of the network and the stream rate.

We have also shown that the size of the probe set affects performance of diffusion schemes, and, in particular, a probe set larger than the actual number of concurrent connections may improve miss ratio/delay performance by modifying the suitable chunk size ranges.

\paragraph{Acknowledgments} This work has been supported by the Collaborative Research Contract Mardi II between INRIA and Orange Labs, and by the European Commission through the NAPA-WINE Project, ICT Call 1 FP7-ICT-2007-1, Grant Agreement no.: 214412.
\newpage
\bibliographystyle{abbrv}
\bibliography{RR-7032}

\end{document}